\documentclass{aa}  
\usepackage{graphicx}
\usepackage[usenames]{color}
\begin{document}
\title{Evolved stars hint to an external origin of  enhanced metallicity in planet-hosting stars}


   \author{
   	L. Pasquini\inst{1}
	\and
	M.P. D\"ollinger\inst{1,4}
          \and
          A. Weiss\inst{4}
          \and 
   	L. Girardi\inst{3}
	      \and
          C. Chavero\inst{5,6} 
	\and       
	A. P. Hatzes\inst{2}
   	 \and
	   L. da Silva \inst{6}
	   \and 
	   J. Setiawan\inst{7}
	  }
	  
   \offprints{
    Luca Pasquini  \email{lpasquin@eso.org}
        }

     \institute{
European Southern Observatory, Garching bei M\"unchen, Germany
         \and
	Th\"uringer Landessternwarte Tautenburg,
                Sternwarte 5, D-07778 Tautenburg, Germany
		\and INAF-Osservatorio Astronomico di Padova, Vicolo dell'Osservatorio 5, I-35122
Padova, Italy
          \and
Max-Planck-Institut f\"ur Astrophysik, Garching bei M\"unchen, Germany
          \and
          Instituto de Astrof\'isica de Canarias, 38200, La Laguna, Tenerife, Spain
	\and
	Observat\'orio Nacional/MCT, 20921-400, Rio de Janeiro, Brasil 
          \and
          Max Planck Institute F\"ur Astronomie, Heidelberg, Germany}
   \date{Received; accepted}

\abstract
  {}
  { 
 Exo-planets are preferentially found around high metallicity main sequence stars. We aim at 
investigating  whether evolved stars share this  
property, and what  this tells  about planet formation.
   }
   {
Statistical tools and the basic concepts of stellar evolution theory
are applied to published results as well as our own 
radial velocity and chemical analyses of evolved stars.
   }
   {
We  show that the metal distributions of planet-hosting (P-H) dwarfs and giants 
are different, and that  the latter do not favor metal-rich systems. 
Rather, these stars follow
 the same age-metalicity relation as the giants without planets in our sample. 
 The straightforward 
explanation is to attribute the difference 
between dwarfs and giants to the much  larger 
masses of giants' convective envelopes. 
If the metal excess on the main sequence is due to pollution,  the effects of dilution
naturally explains why it is not observed among evolved stars. 
}
 { Although we cannot exclude other explanations, the lack of any
   preference for
 metal-rich systems among P-H giants could be 
   a strong indication of the accretion of metal-rich material. We discuss   further 
tests, as well as some predictions and consequences of this hypothesis. 
}

\titlerunning{Accretion in planet-hosting stars}
\maketitle

%
\section{Introduction}
Just a few years after the discovery of the first extra-solar planet (Mayor and Queloz 1995)
 it has become evident that giant planets   are preferentially found around  
 metal-rich solar type stars  (e.g. Gonzalez 1997, 1998, 2001;  Santos et al. 2000, 2001, 2003, 2004,  Fischer \& Valenti 2005). 
Subsequent studies have shown that this preference is real (e.g. not produced by spurious selection effects),  
and that planet-hosting (P-H) stars are on average about 0.25 dex more metal-rich than 
 their counterparts (Santos et al. 2004, 2005).
The immediate question, which is very relevant for the understanding of planet formation, is  
if  this increased metallicity enhances planet formation, or if the high metallicity  
is caused instead by the presence of a planetary system. 

In the first case, favored by the core-accretion  scenario (Pollack et al. 1996), the stars
 should be overmetallic down to their center. This scenario proposes that a 
solid core grows via the accretion of 
planetesimals until it has sufficient mass to capture gas 
from the nebula to form an envelope.
In this case, the planet formation depends strongly on  dust content (Ida \& Lin 2004).

In the second case the star was  polluted by the debris of the planetary system and 
only the external layers were affected by this pollution (Laughlin \& Adams 1997).
This scenario is compatible with 
the gravitational instability mechanism: a gravitationally unstable region in a protoplanetary 
disk forms self-gravitating clumps of gas and dust within which the dust grains coagulate 
and sediment to form a
central core  (Boss 1997). Boss (2002) argues that the gravitational instability model should  
depend very weakly on metallicity, contrary to what is expected from the core accretion scenario.

The pollution scenario has been considered previously and 
several authors have investigated the dependence of 
metallicity on the effective temperature of P-H stars. Since the 
 depth of the convective zone  increases with lower mass along the main sequence,  
 a trend with stellar temperature is expected.  The effect, if present at all,
  is very small and this led most  authors to conclude that  an
  enhanced primordial  
 metallicity of the host stars is favored (e.g. Santos 2005).
  The situation is however complicated because of additional mixing beyond the formal
 convective boundary, either due to thermohaline convection and
 ``metallic fingers'' (Vauclair 2004), or due to other effects which
 manifest themselves in the lithium dip in open clusters (Murray et
 al.\ 2001). 
 
In the last years a few groups started  surveys of evolved stars, G and K giants, with the aim of 
 learning how planets form among more massive stars and of understanding the radial velocity 
variability of these objects (Setiawan et al. 2003b, 2004, Sato et al. 2007 ). 
These surveys along with other sporadic observations led 
to the discovery of 10  exo-planets around 
giant stars. Several authors 
 pointed out that planet-hosting giants are not  preferentially metal-rich systems.   
 The data published were  so few and sparse, on the other
 hand, to prevent any further analysis.

\section{Giants hosting planets}
The 10 G and K  P-H giants 
include  HD~137759 (Frink et al. 2002), HD~47536 and HD~122430 (Setiawan et al. 2003a,b), 
 HD~104985 (Sato et al. 2003), HD~222404(Hatzes et al. 2003), HD~11977 (Setiawan et al. 2005), HD~13189 (Hatzes et al. 2005), $\beta$~Gem (Hatzes et al. 2006), 
 4~UMa (D\"ollinger et al. 2007a), HD~28305 (Sato et al.  2007).
One of the latest:  4~UMa,  is  part of a survey started 4 years ago at the Tautenburg 
observatory  (D\"ollinger et al. 2007a,b,c);  the 
details about the observations, analysis,  and results can be found in the cited papers. 
In addition to  4~UMa,  at least 4 other candidates are 
 present in the Tautenburg survey (D\"ollinger et al. 2007c); adding these to  the stars in literature
 we have a total of 14 giants,  for 10 of  which all parameters have been derived in a 
  homogeneous way.
 
The  metallicity distribution for these 14 giants is  shown in Figure 1 together with  the distribution of
 the P-H main sequence stars  (small black dots; data from the Schneider catalogue). 
 The two distributions have a similar shape, 
 but the giant distribution is  shifted by about 0.3 dex towards lower metallicity. 
A K-S test shows that the probability for both distributions belonging 
 to the same population is around 10$^{-4}$.  Systematic differences,
 such as due
 to the different metallicity scale adopted might be present,  but  they are 
 certainly much  smaller than the shift observed.
We  caution, however, about two points. 

1)  The giant survey is not explicitly biased towards metal-rich stars,
 while the search for planet around  main sequence stars might be (Fischer et al. 2005). 
 Santos et al. (2004) and Fischer \& Valenti (2005) consider that  
 the  shift of  0.25 dex between P-H and not P-H dwarfs is real.
Conversely,  the distribution of P-H giants  follows the general  
 distribution for giants as seen in Figure 2, which shows 
 the age-metallicity distribution of the 
130 giants analyzed by our group (da Silva et al. 2006; D\"ollinger et al. 2007b). 
 
2)  Giants do not posses short period planets, which would
  eventually have been swallowed  by the  expanding stellar envelope. 
  Indeed,  the shortest period 
  for a planet around a giant is 198 days for HD104985 (Sato et al. 2003).  
 In Figure 1 the dashed  line represents  the distribution of the main sequence 
 P-H stars with periods longer than 180 days. 
Clearly, by selecting only the long period planets  the metal distribution of the dwarfs does not change. 
We therefore conclude that  there is a real difference in the metal distribution of 
 main sequence and  evolved P-H stars.  
 Evolved P-H stars are 0.2-0.3~dex more metal-poor than main sequence P-H stars. 
 Interestingly,  this difference is similar to that observed  between P-H and non-P-H main sequence stars.

\begin{figure}[h]
\resizebox{\hsize}{!}{\includegraphics{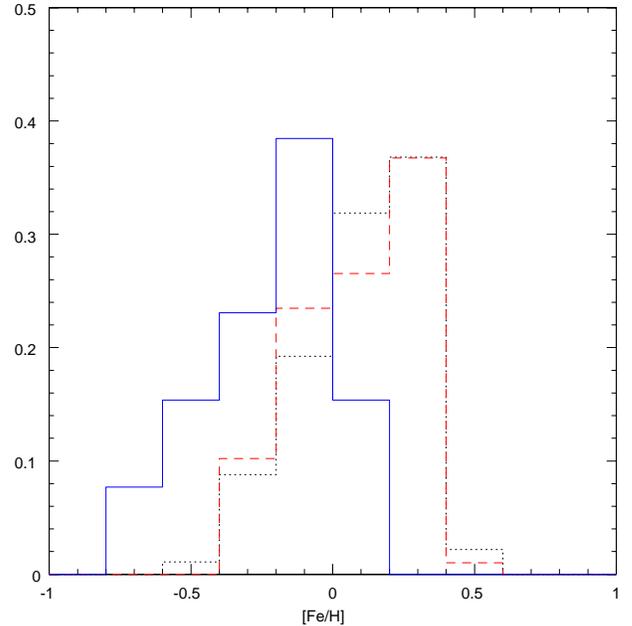}}
\caption{Metal distribution for planet-hosting (P-H) giants (full
  line), P-H dwarfs with periods larger than 180 days (dashed line)
  and all P-H dwarfs (dotted). The giants show a distribution  shifted
  to lower metallicity by about 0.2-0.3~dex with respect to the
  dwarfs.}
\label{fig1}
\end{figure}

\begin{figure}[h]
\resizebox{\hsize}{!}{\includegraphics{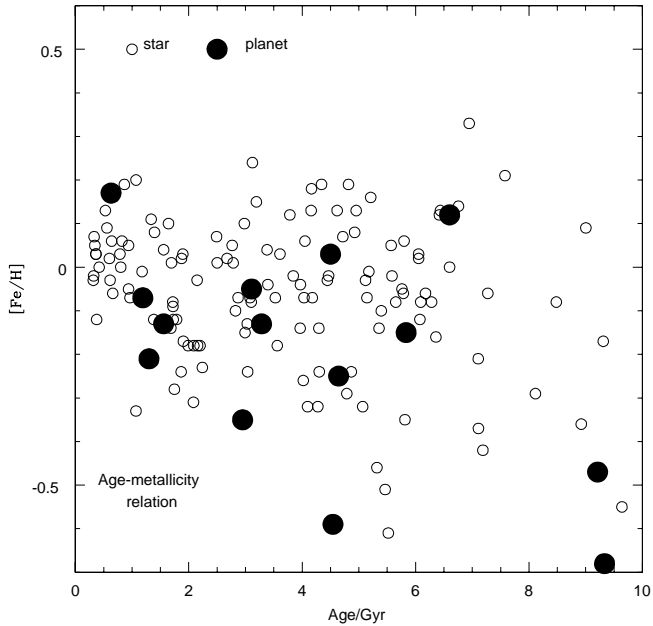}}
\caption{Age-metallicity relationship for all giants belonging to the da Silva et al. (2006) and D\"ollinger et al. (2007b) samples and for the P-H giants, which  closely follow the other  stars. }
\label{fig2}
\end{figure}

\section{Interpretation }

We can think of at least three main differences between main sequence
stars and giants:

1) On average giants have  a somewhat higher mass than the main sequence stars surveyed for planet search.  
Even considering that  the determination of mass in giants suffers  large uncertainties, 
the masses of planet-hosting giants  vary between $\sim$ 0.9  and $\sim$3 M$_{\odot}$
(Da Silva et al. 2006, D\"ollinger et al. 2007b), while  those of main sequence stars are 
between $\sim$ 0.75  and $\sim$ 1.5 M$_{\odot}$. 
Since the fraction of P-H giants  is basically independent of
metallicity,  it is feasible that
intermediate mass stars  favor a planet formation mechanism, such as
gravitational instability, which is independent of metallicity. One could speculate that such a 
mechanism is more efficient in more massive stars, which (likely) have more massive disks.  
 The dependence of metallicity on stellar  mass among main sequence P-H 
 stars  has already been investigated by  Fischer and Valenti (2005) 
 who derived fits to the mass-metallicity distribution of P-H and not P-H stars 
of their sample. Both fits have the same slope and, independent of stellar mass, 
P-H stars are more metal-rich than stars without planets. 
   This result would therefore argue against the  hypothesis that the 
  planet formation mechanism changes significantly with the stellar
  mass. We notice, however,  that the 
  mass range covered by the Fischer and Valenti  study is limited to the 0.8-1.2  M$_{\odot}$, 
  while  by observing giants we cover a larger range of stellar mass.

2)  Giants have on average radii which are about 10 times larger than
 solar-type stars.  If high metallicity favors the migration of planets towards 
short period systems,  metal-rich stars have 
more short period planets than  metal-poor stars. Since  
these planets are  swallowed by the evolved star due to its
large radius, those  stars would be classified as P-H   on the main sequence, 
and non-P-H when evolved. 

The case of  metal-dependent migration has been discussed, for 
example,  in Santos et al.( 2006): higher metallicity should result in a  shorter timescale 
for inward migration. How effective this mechanism could be,  is however a matter of debate;  
Livio and Pringle (2003) find that a decrease in metallicity by a
factor of 10 could slow down by at most a factor 2
the timescale for migration. While these
authors consider this negligible, 
Boss (2005) argues that this factor  is enough to influence the observed 
trend between metallicity and P-H stars. 

The metallicity distribution of Figure 1 is  very different   for  P-H giants and  dwarfs  with  
comparable long orbital periods. This indicates that the 
effects of migration, if present, cannot explain 
the different  metal distribution of P-H dwarfs and giants. 
  
We cannot exclude that  several mechanisms are at work
simultaneously and that they combine to produce the observed distribution. 
A dual formation scheme (one metal dependent, a second   metal independent) has been already  proposed  
(see e.g.  Matsuo et al. 2007).
The  metal-independent planet formation mechanism could be  more effective for larger stellar  masses 
and act therefore on giants much more than on main sequence stars.  
 
3)  The most likely  explanation is  related to the quantity which varies most 
  between dwarfs and giants: the mass  of the convective zone.  
  While in the Sun the fraction of the solar mass
 in the convective envelope $M_\mathrm{ce}$ is  $\sim$0.022
  M$_{\odot}$ ($\log M_\mathrm{ce} = -1.67$) , when the star reaches
 its maximum depth along the RGB, this fraction is about 35 times higher, 
 or almost 0.77 M$_{\odot}$  ($\log M_\mathrm{ce} = -0.11$). 
 In general,  when a 1M$_{\odot}$ star becomes a K giant,  its convective envelope
  is  of the order of 0.7~$M_\odot$. 
 If the high metallicity observed among main sequence stars was confined to the superficial layers,
 with a deepening convective envelope, this would be easily decreased to the
primordial abundance of the star.
 In Fig.~3 we show the fractional mass (in logarithmic units) contained in  the convective
 envelope of stars between 0.8 and 1.5~$M_\odot$, both on the main
 sequence and on the red giant branch where the convective envelope has
 reached its deepest. This indicates the maximum dilution factor. 
Considering, for instance, an excess of  0.25 dex in [Fe/H] (Santos et
al. 2005) in a solar star,  that  would become less than 1\% in   a giant star,
a quantity which is beyond the actual detection capabilities in most
observational cases.  

\begin{figure}[]
\resizebox{\hsize}{!}{\includegraphics[angle=-90,scale=0.4]{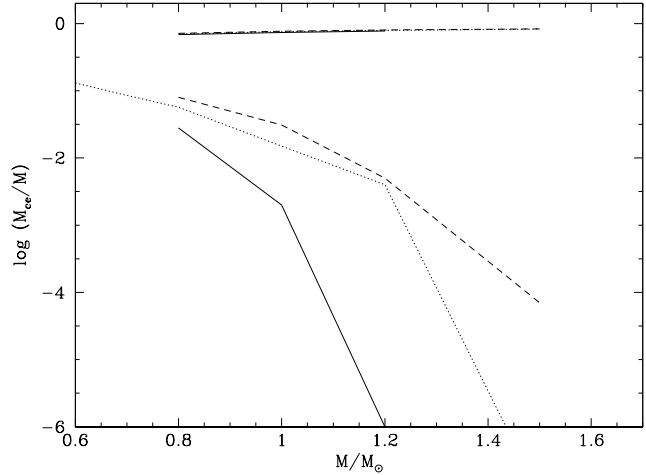}}
\caption{Convective envelope mass $M_\mathrm{ce}$ (logarithmic
  scale; in units of stellar mass) 
for stars of varying mass and in two evolutionary phases.
Shown are three metallicity cases: $Z=0.005$ (solid lines), $Z=0.017$
  (dotted), and $Z=0.026$ (dashed). The lower set of lines corresponds
  to main sequence, the upper to red giant models. 
The drastic enlargement of the
convective mass during the RGB ascent is clearly visible. It is close
to a factor $\sim$50 for a solar-type star.}
\label{fig3}
\end{figure}

As summarized in the introduction,  pollution  has been subject to several investigations
 (Santos et  al. 2005, Ecuvillon et al. 2006, Desidera et al. 2004 for binary systems), which 
 did not find any evidence  for it.  
Previous investigations, however,  were restricted to a limited range 
 of  convective masses, while with giants we greatly enlarge this range. 
 We also emphasize that the actual amount of envelope mass into
  which the accreted metals will be mixed on the main sequence
  is most likely not just the
  convective envelope, but the region into which thermohaline mixing
  reaches (Vauclair 2004), and this is determined mainly by molecular
  weight and not by the stellar mass. 
 We have tested this effect on our own stellar models in the range of 
0.6 to 0.9 M$_{\odot}$, finding that the total mass external to a layer
 with a given molecular weight is 
the same (within  10$\%$), independent  of the stellar mass.  
 If this mixing  is at work, a correlation between stellar metallicity and
 mass (or position on the main sequence) can therefore not be expected.
  The depleted solar Li abundance (M\"uller et al.  1975)  
  clearly shows that the sun  has suffered some form of extra
  mixing, as did stars in open clusters. The additional mass affected
  might be as much as $0.05\,M_\odot$, which is more than twice the present 
  convective envelope mass, but still small if compared to the envelope mass of a giant.  

\begin{figure}[h]
\resizebox{\hsize}{!}{\includegraphics{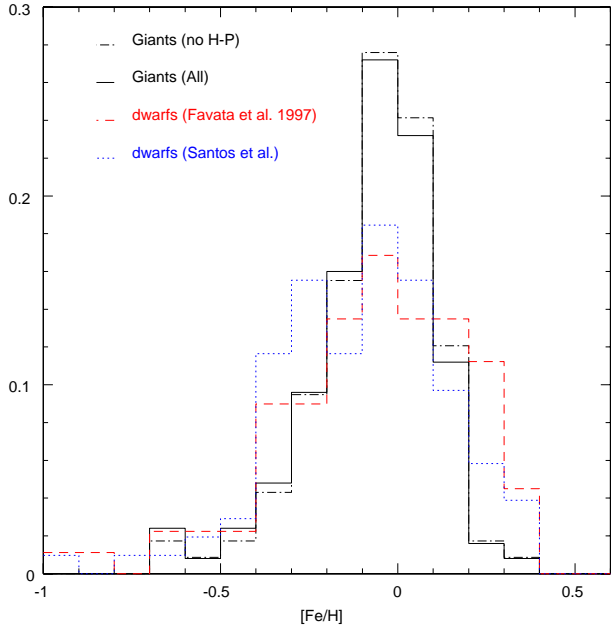}}
\caption{Metal distribution for all the giants from the da Silva et al. (2006) and D\"ollinger et al. (2007b)  
sample  for all stars  (full line) and for not P-H giants (dashed-dotted line),
 for the volume limited sample of Favata et al. (1997, dashed) and Santos et al. (2004, 2005, points). 
An excess of metal-rich stars might be present among the dwarfs. To make such a comparison significant, a number of 
effects in the sample selection and in the analysis should be
considered.}
\label{fig4}
\end{figure}

Even if the observations of giants support the hypothesis that pollution is 
very important, more evidence is required to prove it. A number of 
 interesting tests could be performed.  We need  well controlled samples, 
 in  age, mass,  and internal structure (e.g. if diffusion is at work). 
If the hypothesis of pollution is true, 
we expect an excess of metal-rich stars among main sequence stars  
with respect to an {\it equivalent}  sample of giants.
In Figure \ref{fig4} we show the metal distribution of the giants from 
da Silva et al. (2006) and D\"ollinger et al. 
 (2007b) compared to the distribution of a volume limited sample of 
 main sequence stars from Favata et al. (1997) and  Santos et al. (2004, 2005). 
 As far as the giants  are concerned, we plot  the distribution of  the whole sample and 
the distribution of the non-P-H stars separately. 
No real difference can be discerned  between the two giant distributions. 
 The giants and dwarfs distributions are also very similar, with the main sequence stars showing
 a (not significant)  excess in the highest metallicity bins.
  The comparison between the giants and the Favata et al. (1997) results in particular 
 suggests that  the small  excess of metal-rich dwarfs is almost perfectly compensated by an 
 excess of solar-metallicity giants, which is exactly the signature we would expect from  pollution. 
However this excess is mostly 
due to the coolest main sequence stars and  other aspects, such as age distribution and 
galactic evaporation, should be taken into account to properly compare the data (Favata et al. 1997). 
 
Any difference in the correlation between the presence of
 planets and metallicity should also become evident when
 observing stars on the hot and cool parts of the SGB, where
 the convective zone deepens by 
 a factor 10-100 in a relatively short interval of magnitude and time. 
 Fischer and Valenti (2005) 
 searched for a slope in the metallicity distribution of 
 subgiants but  did not find  any. Murray et al. (2001), however,  did find evidence for 
 lower metallicity in  Hertzspung gap stars with respect to their main sequence sample. 
 A dedicated study should be devoted to this point. 

Open clusters and associations might be optimal sites for investigating the effects of pollution. 
In the presence of P-H stars, a  direct measurement  should reveal 
an excess of metallicity, or at least a larger spread  among the main sequence stars, 
but not among the giants belonging to the same cluster.
This investigation could be also extended to the cooler part of the main sequence where the 
 convective zone is  significantly deeper than for solar-type stars.  Most interesting could be a  search  for `outliers':
 proper motions and/or radial velocity open clusters' members 
 with discrepant (higher) metallicity. This  could be an efficient 
 way of identifying  P-H candidates 
 in open clusters and associations and to prove  the pollution hypothesis. 

\section{Conclusions}
 By enlarging the number of giants hosting exoplanets, it has been possible to establish that 
 their metallicity distribution is very different from that of planet-hosting
 main sequence stars.
Giants hosting exoplanets do not favor high metallicity objects, but
 follow the age-metallicity distribution observed for all stars surveyed. 
 
 The interpretation of the data is not straightforward: a scenario which 
 includes strong differences in planet formation  with stellar  mass
   and possibly planet migration 
 is plausible, but the most immediate explanation  is that the high metallicity observed among main 
  sequence stars is  due to pollution of their atmospheres. The metal excess produced by this pollution,
 while visible in the thin atmospheres of solar-like stars, it  is 
completely diluted in the extended, massive envelopes of the giants. 
This interpretation is in apparent contrast with results on main sequence stars 
obtained by several groups (Fischer and Valenti 2005, Ecuvillon et al. 2006 among others), 
which favor the primordial scenario, where stars are born in high metallicity clouds.
We believe that the possible explanation of this discrepancy is that the effects of 
pollution are rather tiny on the main sequence and difficult to 
 detect. The fact that 
Fischer and Valenti(2005)  do not find any evidence for dilution among
 a subsample of subgiants is of greater concern.
The subgiants analyzed by Fischer and Valenti (2005) belong to  the  survey of 
156 subgiants (evolved A stars) by Johnson et al. (2006, 2007).
To the best of our knowledge, this planet - search survey is still on going.  
 The four planets published from this survey, with metallicity of (Fe/H=0.11, -0.15, 0.12 and -0.07, Johnson et al. 2006, 2007) 
 are very compatible with a normal metal distribution.
 We are eager to  see the final results of this and similar surveys and to compare their metal distributions
 with our.


\begin{acknowledgements}
Partially based on observations made with the 2-m Alfred Jensch Telescope of T-L Tautenburg. 
M.D. was supported by ESO DGDF. 
The Schneider planet encyclopedia has been used (http://vo.obspm.fr/exoplanetes/encyclo/index.php).
\end{acknowledgements}

\end{document}